\documentclass[10pt,a4paper]{article}

\normalsize \textwidth 454pt \textheight 690pt \baselineskip 14.5pt \columnsep 12pt \columnwidth
230pt \topskip 45pt \evensidemargin -26pt \oddsidemargin 9pt \topmargin -15 mm
\usepackage{graphicx, natbib, amssymb}

\usepackage[centertags]{amsmath}

\numberwithin{equation}{section}

\begin{document}

\title{On the $q-$parameter spectrum of generalized information-entropy measures with no cut-off prescriptions}

\author{Marco Masi\footnote{Corresponding author: marco\!\_masi2@tin.it, fax: +39 02 6596997} \\ \small via Tiepolo 36, 35100, Padova, Italy.}

\maketitle

\begin{abstract}
After studying some properties of the generalized exponential and logarithmic function, in
particular investigating the domain where the first maintains itself real and positive, and
outlining how the known dualities $q \leftrightarrow \frac{1}{q}$ and $q \leftrightarrow 2-q$ play
an important role, we shall examine the set of q-deforming parameters that allow generalized
canonical maximum entropy probability distributions (MEPDs) to maintain itself positive and real
without cut-off prescriptions. We determine the set of q-deforming parameters for which a
generalized statistics with discrete but unbound energy states is possible.
\end{abstract}

\vspace{5mm}

entropy principle, duality, cut-off prescriptions \\


\newpage

\section{Introduction}

MEPDs which emerge from generalized measures, like Tsallis \cite{Tsallis} and Renyi \cite{Renyi}
entropies, subjected to Jaynes' maximum entropy principle, i.e. constrained by a q-generalized (or
escort) energy mean value and the normalization condition, are expressed in terms of q-deformed
generalized exponential functions of the energy microstates (also called "q-maxent
distributions"). If these MEPDs are supposed to have a physical and statistical meaning, then they
must be real and positive. The q-exponential function however is not always a real and positive
quantity. Usually, to avoid negative or complex values of the microstate probabilities for some
energy values, one introduces cut-off prescriptions. One example of this is the "Tsallis cut-off
prescription" \cite{Curado} which simply sets $p_{i}=0$ whenever this is the case. This is not
necessarily an ad-hoc requirement. For instance in self-gravitating systems, as in the case of
stellar polytropic distributions, cut-offs arise naturally and correspond to a gravitational
escape velocity \cite{Plastino}. Another scenario leading naturally to Tsallis' cut-off is the
case of a system S with energy levels $\varepsilon_{i}$ coupled to a finite heat-bath B with
quasi-continuous energy level distributions. A. R. Plastino and A. Plastino showed
\cite{plasandplas} that if the heat-bath's number of states with energy less or equal than, say E,
increases as a power law of E, then Tsallis' cut-off implies the obvious fact that the probability
to find the system with energy $\varepsilon_{i} \geq E_{0}$, with $E_{0}$ the total energy of the
"total" system S+B, is zero.

However, there are other systems where one has no reason to set a priori a cut-off on the energy
levels (e.g. on asymptotic behaviors like in a Maxwellian velocity distribution or a black body
radiation curve or other quantum distributions). For these kind of situations Teweldeberhan,
Plastino and Miller \cite{Teweldeberhan} suggested to redefine the q-generalized exponential
function itself such that "\textit{it leads, via the maximum entropy principle, to quantum
distribution functions that have been very successful in the study of a concrete and important
physical phenomena: high $T_{C}$ superconductivity}", and working with this special case the
thermodynamical consistency of their cut-off prescription is shown. Proposing alternative
definitions of the q-exponential function as has done in this case is one possible way and is
inescapable if one wants to maintain a continuously varying $q$-parameter. However, we might also
wonder which are the discrete values of the $q$-parameter for which the generalized exponential
remains nevertheless always real and positive such that no cut-off and no redefinition of the
q-exponential function is needed.

Here we will investigate for this purpose the deformed logarithm-exponential representation of
probability distributions (PD) and entropy measures subjected to Jaynes' MEP focussing our
attention on the dual description of generalized statistics. We will first study some properties
of the generalized exponential and logarithmic function and, investigating the domain where the
former maintains itself real and positive, outline what role the known dualities $q
\leftrightarrow \frac{1}{q}$ and $q \leftrightarrow 2-q$ play. This will then be applied to
microcanonical PDs and $q$-deformed generalized entropy measures. We will show how these
q-generalized entropy measures and MEPDs with no cut-off prescriptions (i.e. no energy
constraints) are positive and real (as any meaningful physical measure should be) only for
discrete values of the $q$-parameter and will determine these explicitly.

\section{\label{secspectrum} The spectrum of the \lowercase{q}-deforming parameter for real generalized exponentials}

Throughout the literature the application of the $q$-deformed generalized exponential and
logarithm function has become commonplace. As it is well known the usual Euler exponential $e^{x}$
can be defined as the solution of the differential equation $y(x)'=y(x)$. The q-deformed
exponential can be regarded as a generalization in the sense that it can be obtained from the
solution of \begin{equation}y'(x) = y(x)^{q}\label{q-expd}\end{equation} and initial condition
$y(0)=1$, with $q$ a real parameter, as
\begin{equation}e_{q}^{x} \equiv e_{q}[x]=[1+(1-q)x]^{\frac{1}{1-q}} \, ,\label{q-exp} \end{equation}
which becomes the exponential function for $q=1$. The inverse of the q-exponential is the
\textit{generalized q-logarithm function} \[ \log_{q} x = \frac{x^{1-q}-1}{1-q} \, , \] which, for
$q=1$, becomes the common Napier logarithm.

Useful for our purposes will be the fact that \begin{equation} e_{q} [x+y] = e_{q} [\gamma x] \,
e_{q} [y] \, \, ; \hspace{15mm} \forall x,y \in \mathbb{R} \label{q-shift}\end{equation} with
$$\gamma= \frac{1}{1+(1-q)y} \,.$$

There are many indications \cite{Tsallis2} which suggest that statistical theories based on
functionals of power law PDs (or probability density functions) obeying (\ref{q-expd}) describe
correctly several complex systems.

Now, the generalized exponential (\ref{q-exp}) is a quantity of type $e_{q}[x] = a(x,q)^{b(q)}$
with
\[a(x,q)=1+(1-q)x \hspace{5mm} ; \hspace{5mm} b(q)=\frac{1}{1-q} \, ,\] and which can take real or
complex, positive or negative values according to what we choose for $a(x,q)$ and $b(q)$.
Obviously, if $a(x,q)$ is a real and positive quantity for some $x$, then $a(x,q)^{b(q)}$ is
always real for every $b(q) \in \mathbb{R}$, i.e. for every $q$ such that
$e_{q}[x]=[1+(1-q)x]^{\frac{1}{1-q}}$ remains real and positive. But once an interval
$I_{x}=[x_{1},x_{2}]$ for $x_{1}, x_{2} \in \mathbb{R}$ is chosen, also a continuous interval
$I_{q}^{cont}=[q_{1},q_{2}]$ is fixed so that $a(x,q)=1+(1-q)x \geq 0$ always. While for every $q$
outside $I_{q}^{cont}$ it is $a(x,q)<0$, $\forall x \in I_{x}$. In this case $e_{q}[x]$ might
become complex. To know where this precisely does \textit{not} occur we can write
\[e_{q}[x]=a(x,q)^{b(q)}= (-1)^{b(q)} |a(x,q)|^{b(q)} = e^{i\pi b(q)} |a(x,q)|^{b(q)} \, ,\] which
is real if and only if $b(q)=k$ for every $k \in \mathbb{Z}$ (and is positive for $k$ even and
negative for $k$ odd).

Then, defining \[e_{q(k)}[x]= \left(1+\frac{x}{k}\right)^{k} \, \; \hspace{6mm} k \in
\mathbb{Z}/\{0\}\vspace{2mm}\] as the ``\textit{discrete spectrum generalized exponential}'', to
distinguish it from the ``\textit{continuous spectrum generalized exponential}'' $e_{q}[x]$, one
establishes that the generalized exponential is real on all $I_{x}$ iff $b(q)=\frac{1}{1-q}=k \in
\mathbb{Z}$, i.e. when $e_{q}[x] \equiv e_{q(k)}[x]$ with
\[\left\{q(k)\right\}=\left\{\frac{k-1}{k} ; \, k \in \mathbb{Z} / \{0\} \right\} = \left\{0,
\frac{1}{2}, \frac{2}{3}, \frac{3}{4}, ... , 1\right\} \cap \left\{1, ... , \frac{4}{3},
\frac{3}{2}, 2\right\} \, .\]

The figures in the appendix trace the behavior of the discrete spectrum generalized exponential
for some examples of the above set of $q$-parameters.

The q-exponential is always positive for $k$ even (Fig.1 with $k \in \mathbb{Z^{+}}/\{0\}$, and
Fig.2 with $k\in \mathbb{Z^{-}}/\{0\}$), i.e. given $l\in \mathbb{Z}/\{0\}$ for all
\[\left\{q(l)\right\}=\left\{\frac{2l-1}{2l}\right\} = \left\{\frac{1}{2}, \frac{3}{4},
\frac{5}{6}, ... ,1\right\} \cap \left\{1, ... , \frac{7}{6}, \frac{5}{4},
\frac{3}{2}\right\}\,.\] It can become negative if $k$ is odd (Fig.3 with $k \in
\mathbb{Z^{+}}/\{0\}$, and Fig.4 with $k \in \mathbb{Z^{-}}/\{0\}$), i.e. given $l\in \mathbb{Z}$
for all \[\left\{q(l)\right\}=\left\{\frac{2l}{2l+1}\right\} = \left\{\frac{2l}{2l-1}\right\} =
\left\{0\right\} \cap \left\{\frac{2}{3}, \frac{4}{5}, \frac{6}{7} ... ,1\right\} \cap \left\{1,
..., \frac{6}{5}, \frac{4}{3},2\right\} \, ,\] where $q=1$ is a point of accumulation of the
discrete spectrum.\footnote{From the graphs it becomes also clear how to approximate $e_{q}[x]
\approx e^{x}$ for $q \approx 1$, must be done cautiously since $e_{q}[x]$ can diverge in some
regions from Euler's exponential for every value which is not exactly $q=1$. On the other side, if
PD are expressed with q-exponentials, the infinities for $q>1$ and $x>0$ (in Fig. 2 and Fig.4)
might at first look worrisome. However, as we will see later, the partition function of MEPDs
renormalizes these divergences (see eq. \ref{pitotal2}). The discontinuities of the q-exponential
should cause no concern for stationary \textit{normalized} PDs.}

Therefore, in general, outside the continuous parameter space for $q$, in order to maintain a
generalized exponential to be a real quantity on all the domain $I_{x}$, only discrete values of
$0 \leq q \leq 2$ are allowed.

Now, note also how there exists (except, of course, for the special case $q=0$) a one-to-one
correspondence \begin{equation}\pi: q \leftrightarrow \frac{1}{q}\label{qpai}\end{equation} (i.e.
the Fig.1$\leftrightarrow$Fig.4 and Fig.2$\leftrightarrow$Fig.3 correspondences) of the discrete
spectrum between the real even (always positive) and a real odd (negative somewhere) generalized
exponential function. Also a \begin{equation}\psi: q \leftrightarrow 2-q
\label{qpsi}\end{equation} correspondence between the two always positive (Fig.1 $\leftrightarrow$
Fig.2) and the two sign-changing (Fig.3 $\leftrightarrow$ Fig.4) generalized exponentials is
established. These are reminiscent of the $q \leftrightarrow \frac{1}{q}$ and $q \leftrightarrow
2-q$ dualities first noted by Tsallis, Mendes and Platino \cite{Tsallis4} (and further
investigated by Naudts\cite{Naudts2}, Abe and Okomoto\cite{Abe2}, Frank and Plastino\cite{Frank}
and others), and acquire here a new meaning.

Then it is quite natural to search for the bijections between Fig.1$\leftrightarrow$Fig.3 and
Fig.2$\leftrightarrow$Fig.4 (i.e. between the sign-changing and the always positive generalized
exponentials). These are given by the $\phi: q\rightarrow\frac{1}{2-q}$ correspondence (Fig.3
$\rightarrow$ Fig.1 and Fig.4 $\rightarrow$ Fig.2), and $\varphi: q\rightarrow 2-\frac{1}{q}$
(Fig.1 $\rightarrow$ Fig.3 and Fig.2 $\rightarrow$ Fig.4).

All these are correspondences which have their roots in in the complex coniugate
$\overline{\overline{z}}=z$, $z \in \mathbb{C}$. However, the $\pi$ and $\psi$ transformations are
involutions, while $\phi$ and $\varphi$ are clearly not. $\pi$ and $\psi$ appear to be more
fundamental in the sense that $\phi$ and $\varphi$ are a composition of the first two. One can of
course construct an infinite number of correspondences between the discrete spectra but only as
compositions of these. And another decisive difference is that these represent a transformation
between a $q>1$ and $q<1$ statistics, while $\phi$ and $\varphi$ always map on the same $q<1$ or
$q>1$ statistics.

Putting all this together we can finally say that two possibilities exist.

1) $e_{q}[x]=a(x,q)^{b(q)}=[1+1(1-q)x]^{\frac{1}{1-q}}$ is real and positive on all the domain
interval $I_{x} =[x_{1},x_{2}] \in \mathbb{R}$ for the continuous parameter spectrum $q \in
I^{cont}_{q}=[q_{1},q_{2}]$, where $I_{x}$ and $I_{q}^{cont}$ are determined by the set of

\[q \begin{cases}
\,\,\, \leq 1 + \frac{1}{x} &  \forall \,x\!> 0 \cr

\,\,\, \geq 1 + \frac{1}{x} & \forall \,x\!< 0 \cr\cr

\,\,\,  < + \infty & \text{if $x=0^{+}$} \cr

\,\,\,  > - \infty & \text{if $x=0^{-}$} \cr

\end{cases}
\hspace{10mm} \mathrm{and} \hspace{10mm} x
\begin{cases}
\,\,\, \geq \frac{1}{q-1} & \forall \,q\!< 1 \cr \,\,\,

\leq \frac{1}{q-1} & \forall \,q\!> 0 \cr\cr

\,\,\,  < + \infty & \text{if $q=1^{+}$} \cr

\,\,\,  > - \infty & \text{if $q=1^{-}$} \cr
\end{cases}\]

\vspace{3mm}

2) The generalized exponential can still be real on the entire domain $I_{x}$ also outside
$I^{cont}_{q}$, but only for a discrete parameter spectrum $I^{dis}_{q} = \{q(k)\}$ with
\[q(k)=\frac{k-1}{k} \hspace{2mm} \in \, [0,2] / I^{cont}_{q} \,\, ; \hspace{6mm}  (k \in
\mathbb{Z}/\{0\}) \, ,\] where $e_{q(k)}[x]=(1+\frac{x}{k})^{k}$ remains always positive for $k$
even, but changes sign in $x^{*}=\frac{1}{q-1} = -k$, if $k$ is odd.

Cases of particular interest where one might need to know where $e_{q}[x]$ is non complex on all
$I_{x}$ are for instance

\[I_{x}=]-\infty,0]=\mathbb{R^{-}} \Rightarrow e_{q}[x] \in \mathbb{R}
\Leftrightarrow\] \begin{equation}q \in I_{q}^{cont}= [1, + \infty[ \hspace{5mm} \mathrm{or}
\hspace{5mm} q(k) \in I_{q}^{dis}=\left\{\frac{k-1}{k}; k \in \mathbb{Z^{+}}/\{0\}\right\} \in
[0,1[\label{eqr-}\end{equation}

\[I_{x}=[0, +\infty[=\mathbb{R^{+}} \Rightarrow e_{q}[x] \in \mathbb{R} \Leftrightarrow\]
\begin{equation}\hspace{2mm} q \in I_{q}^{cont}= ]- \infty, 1] \hspace{5mm} \mathrm{or} \hspace{5mm} q(k) \in
I_{q}^{dis}=\left\{\frac{k-1}{k}; k \in \mathbb{Z^{-}}/\{0\}\right\} \in
]1,2]\label{eqr+}\end{equation}

\[I_{x}=]-\infty,+\infty[=\mathbb{R} \Rightarrow e_{q}[x] \in \mathbb{R} \Leftrightarrow\]
\begin{equation}\hspace{3mm}q \in I_{q}^{cont}= \{1\} \hspace{5mm} \mathrm{or} \hspace{5mm} q(k) \in
I_{q}^{dis}=\{\frac{k-1}{k}; k \in \mathbb{Z}/\{0\}\} \in [0,2]/\{1\} \label{eqr-+}\end{equation}

\vspace{6mm}

When the k-values over the discrete spectrum are even, then we are restricting on $e_{q}[x] \in
\mathbb{R^{+}}$.

In general, expanding the dominion $I_{x} \in \mathbb{R}$ for which we require $e_{q}[x]$ to be
real means expanding the discrete spectrum while reducing the continuous one, and viceversa.
Moreover, for every $q \in I_{q}^{dis}$ then $\frac{1}{q} \in I_{q}^{cont}$, but in general not
always the viceversa is true. In any case this holds always for some $q$'s between a $1 \leq q
\leq 2$ statistics and a $\frac{1}{2} \leq q \leq 1$ statistics.

From case (\ref{eqr-+}) we see that if we want a generalized exponential to be real on all the
real dominion only discrete values for $q$ are allowed because the continuous spectrum reduces to
a single element, i.e. $q=1$. With this, one incidentally discovers that the known sequence which
leads to the good old Euler function \[\lim_{k\rightarrow \pm \infty} e_{q(k)}[x] =
\lim_{k\rightarrow \pm \infty} \left(1+\frac{x}{k}\right)^{k} =e^{\pm x} \, ,\] has its origin and
interpretation as the limit of the sequence of discrete spectrum generalized exponentials, i.e.
the limiting case (converging to the accumulation point in $q=1$) of the generalized exponentials
which are real on all $\mathbb{R}$.

So far we were concerned with things from a purely analytic point of view. We can now direct our
attention to generalized exponentials in the frame of entropies and microcanonical probabilities.

\section{The discrete \lowercase{q}-deforming spectrum for real and positive MEPDs and entropies}

In statistical physics entropy can be defined as the logarithm of the phase space volume
($\Gamma$-space) of the entire system \[S=\log\Gamma\ \, . \] In order to show that what we are
trying to do has a validity which goes beyond Tsallis' entropy, we keep very general to any form
of entropy measures which lead to q-exponential MEPDs. These are all those measures where one
extends from Napier's logarithm to a second parameter $r$-generalized logarithm function as
follows\footnote{This is of course a somewhat heuristic development for an entropy generalization,
but it is sufficient for our purposes. For a more rigorous explanation of all that see also
\cite{Frank} and \cite{Masi}.}
\begin{equation}S_{SM}= \log_{r} \Gamma(P,q)  = \log_{r} e_{q}^{S_{T}} = \frac{1}{1-r} \left[
\left( \sum_{i} p_{i}^{\,q} \right) ^{\frac{1-r}{1-q}} - 1 \right] \, , \label{eqit}
\end{equation} where \[\Gamma(P,q) = \left(\sum_{i}p_{i}^{q}\right)^{\frac{1}{1-q}} =
e_{q}^{S_{T}}\, , \] and \[S_{T} = \frac{\sum_{i}p_{i}^{q}-1}{1-q} \, , \] is Tsallis' entropy
\cite{Tsallis}. For $r=q$ one obtains Tsallis entropy, if $r=1$ we have \[S_{R} = \log \Gamma(P,q)
= \frac{1}{1-q}\log \sum_{i} p_{i}^{q} \, , \] which is Renyi's entropy \cite{Renyi}, while in
general \ref{eqit} can be recognized as the less known entropy of B.D Sharma and D.P Mittal
\cite{Sharma}. It can be shown that \ref{eqit} is the most general pseudo-additive measure which
includes Tsallis' and Renyi entropies based on escort averages statistics admitting of a partition
function(\cite{Daffertshofer}, \cite{Frank}). This "$\log_{r} \times$ exp$_{q}$"-form
representation will later make it easier to recognize the $q$-spectrum of $\Gamma(P,q)$.

Now, one can check \cite{Tsallis} with the method of the Lagrangian multipliers that, applying
Jaynes' maximum entropy principle, the generalized MEPDs are (lets label with $\widehat{p}_{i}$
the stationary PDs of family $\widehat{P}$)\footnote{We did obtain the MEPDs not only for Tsallis'
entropy, as indicated in the reference, but for a more general set of entropies. The analytic
procedure of maximization is of course equivalent.}\begin{equation}\widehat{p}_{i} =
\frac{e_{q}\left[-\beta_{g}(E_{i}-U)\right]}{Z(\widehat{P})} \, ,\label{pitotal2}\end{equation}
with $$Z(\widehat{P}) = \sum_{j} e_{q} \left[-\beta_{g} (E_{j}-U) \right]$$ a generalized
partition function, where $E_{j}$ is the j-th energy microstate, $U$ the mean energy of the system
and $\beta_{g}$ the generalized Lagrangian parameter (the inverse temperature in BG-statistics).
The $r$-parameter gets absorbed in $\beta_{g}$.

It might be useful to recall that expressions like that in \ref{pitotal2} are invariant under the
choice of the mean energy $U$. Because of \ref{q-shift}, one can rewrite it as
\begin{equation}\widehat{p}_{i} = \frac{e_{q}\left[-\overline{\beta}_{g}E_{i}\right]}{\overline{Z}(\widehat{P})}
\label{pizen}\end{equation} with $$\overline{Z}(\widehat{P}) = \sum_{j} e_{q}
\left[-\overline{\beta}_{g} E_{j} \right] = \frac{Z(\widehat{P})}{e_{q}[\beta_{g} U]}\, ,$$ and
\[\overline{\beta}_{g} = \beta_{g} \, \frac{1}{1 + (1-q) \beta_{g} U} \, .\]

As we have seen (and tried to depict in the figures in the appendix), the transformations of the
deforming parameter (\ref{qpai}) and (\ref{qpsi}) establish a relationship between the discrete
generalized exponentials which are (positively or negatively) real. MEPDs \ref{pitotal2} are in
form of generalized exponentials and must therefore have for every parameter $q$ their real
dominion. (\ref{pizen}) is especially useful to evaluate for which values of $q$ the microstate
probabilities, $\widehat{p}_{i}$, with no energy cut-off, are always real and positive. If we
consider $\overline{\beta}_{g}$ being positive, then we are dealing with the case
(\ref{eqr-})\footnote{The case for $\overline{\beta}_{g}<0$ is \ref{eqr+} and is more delicate
since it implies negative temperatures if one considers the lagrangian parameter proportional to
an inverse temperature. We will not consider the physical implications of this here, but
analytically one proceeds in the same way.} and must conclude that if we want the generalized
optimized PDs to have statistical and physical significance, i.e. to be real and positive on all
$\mathbb{R^{-}}$ ($k$ even), then we are forced to restrict our choice to a continuous parameter
spectrum $I_{q}^{cont} = \left\{q \geq 1 \right\}$ or on a discrete parameter spectrum
\[I_{q}^{dis}= \left\{\frac{2l-1}{2l}; l \in \mathbb{N}/\{0\}\right\} = \left\{\frac{1}{2},
\frac{3}{4}, \frac{5}{6}, ..., 1\right\}\,.\]

On the other side consider the q-generalized $\Gamma$-space of the entropy in its ``$\log_{r}
\times$ exp$_{q}$''-form, $\Gamma=e_{q}^{\,\,S_{T}}$. Also the phase space volume and Tsallis'
entropy are supposed to be a statistical meaningful physical quantity only if they are real and
positive. If no cut-off is imposed on the entropy and since $S_{T} \in \mathbb{R}^{+}$, then case
(\ref{eqr+}) determines the $q-$spectrum as $I_{q}^{cont} = \{q\leq1\}$ and
\[I_{q}^{dis} = \left\{\frac{2l-1}{2l}; l\in \mathbb{Z}^{-}/\{0\}\right\} = \left\{1, ...
,\frac{7}{6}, \frac{5}{4}, \frac{3}{2}\right\} \, .\]

Putting this together, we can say that any physically meaningful generalized q-exponential
statistics based on Jaynes' maximum entropy principle (i.e. with entropies and the associated
MEPD's as real and positive scalars) with no cut-off prescriptions must be restricted to case
(\ref{eqr-+}) and the $q$-deforming spectrum can have only discrete values
\begin{equation}I_{q}^{dis} = \left\{\frac{2l-1}{2l}; l\in \mathbb{Z}/\{0\}\right\} =
\left\{\frac{1}{2}, \frac{3}{4}, \frac{5}{6}, ..., 1, ... ,\frac{7}{6}, \frac{5}{4},
\frac{3}{2}\right\} \, ,\label{finalspectrum}\end{equation} with $q=1$ a point of accumulation.

This was a strictly formal treatment and in the frame of a thermostatistical reasoning this does
of course not show that nature chooses just the values for q of (\ref{finalspectrum}). But if so,
also the energy states of systems described by this statistics must be discrete. Since, in some
physical systems, there are no physical reasons to introduce a priori energy cut-offs, one can
conjecture that these parameters might be preferred over others. This might give also a hint for
further experimental work that tries to highlight if nature indeed prefers to be described by
general exponentials or not. Because the necessity to introduce cut-offs can also suggest that the
generalized exponential function is not the correct, or only an approximate form to describe an
extended statistics. Empiric evidence is to our knowledge so far still affected by too much
incertitude. However, the best experimental and statistical measurement we are aware of might be
the work of Silva, Alcaniz and Lima \cite{Silva} with plasma oscillation data who claim to have
obtained a value of $q=0.77 \pm 0.03$ at 95\% confidence level. This would be in line with our
value $q(k=4)=q(l=2)= \frac{3}{4}\,.$ Of course only further experimental evidence can confirm or
dismiss this hypothesis.

\section{Conclusion}

We studied the q-parameter (continuous and discrete) spectrum on which a generalized exponential
is real and positive and concluded that assuming that Jaynes' maximum entropy principle holds also
in a generalized q-statistics, if we want to have physically meaningful real and positive MEPDs
and entropy scalars on an energy range without cut-off prescriptions, then only discrete
parameters $q \in [\frac{1}{2}, \frac{3}{2}]$ are allowed. This does not mean that a q-statistics
with different values than those given in \ref{finalspectrum} isn't possible, but then the
introduction of limited energy ranges or the redefinition of a new q-generalized exponential is
inescapable.

\newpage

\ \vspace{-10mm}

\begin{appendix}
\setcounter{section}{1}
\begin{center}
\textbf{Appendix}
\end{center}
\vspace{-1mm} \small

\hspace{13mm}\textbf{Graphs and relationships between the discrete generalized exponentials}

\ \vspace{-0mm}

\begin{center}
\includegraphics[scale=0.84]{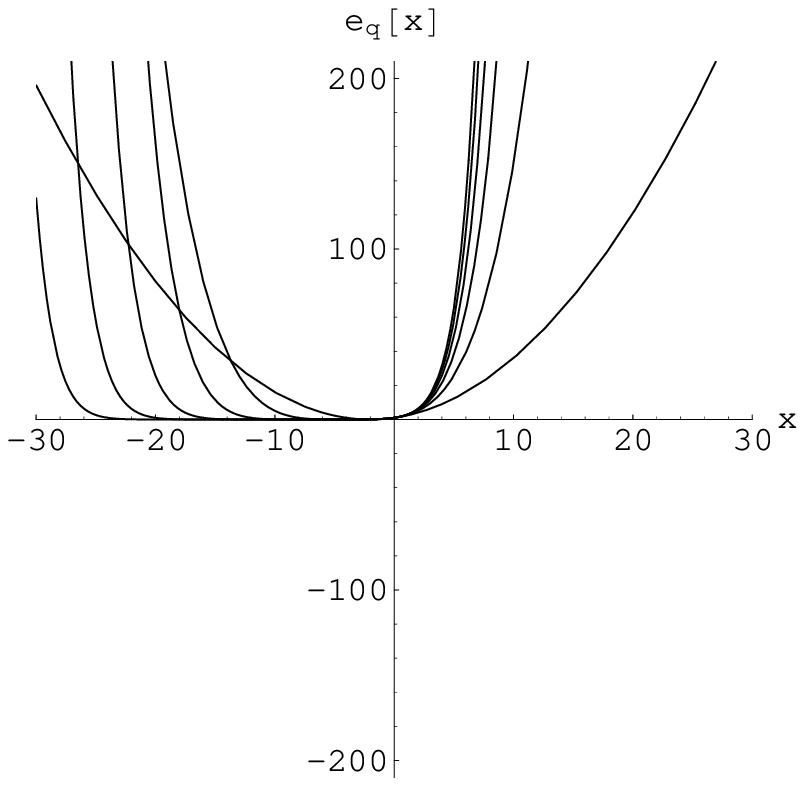}
\includegraphics[scale=0.84]{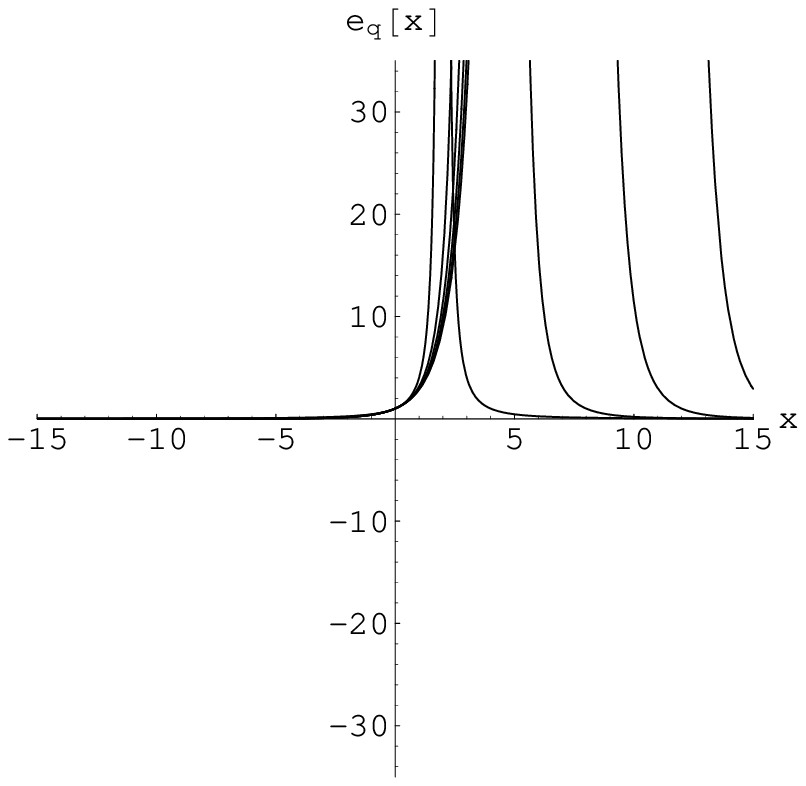}
\end{center}

\hspace{7mm} \footnotesize Fig.1 $\,q=\{\frac{1}{2}, \frac{3}{4}, \frac{5}{6}, \frac{7}{8},
\frac{9}{10}, \frac{11}{12}$\} \hspace{22mm} Fig.2
$q=\{\frac{13}{12},\frac{11}{10},\frac{9}{8},\frac{7}{6},\frac{5}{4}, \frac{3}{2}$\}\normalsize

\vspace{5mm}

\begin{picture}(30,65)(15,0)

\put(112,70){\vector(2,-1){195}} \put(112,70){\vector(-2,1){1}}

\put(308,70){\vector(-2,-1){195}} \put(308,70){\vector(2,1){1}}

\put(200,3){\scriptsize $q \leftrightarrow \frac{1}{q}$}

\put(125,70){\vector(1,0){170}} \put(125,70){\vector(-1,0){1}}

\put(194,60){\scriptsize $q \leftrightarrow 2-q$}

\put(125,-27){\vector(1,0){170}} \put(125,-27){\vector(-1,0){1}}

\put(194,-35){\scriptsize $q \leftrightarrow 2-q$}

\put(115,58){\vector(0,-1){75}}

\put(72,5){\scriptsize $q \rightarrow \frac{1}{2-q}$}

\put(110,-15){\vector(0,1){75}}

\put(120,25){\scriptsize $q \rightarrow 2-\frac{1}{q}$}

\put(305,58){\vector(0,-1){75}}

\put(315,5){\scriptsize $q \rightarrow \frac{1}{2-q}$}

\put(310,-15){\vector(0,1){75}}

\put(265,25){\scriptsize $q \rightarrow 2-\frac{1}{q}$}

\end{picture}

\vspace{10mm}

\begin{center}
\includegraphics[scale=0.84]{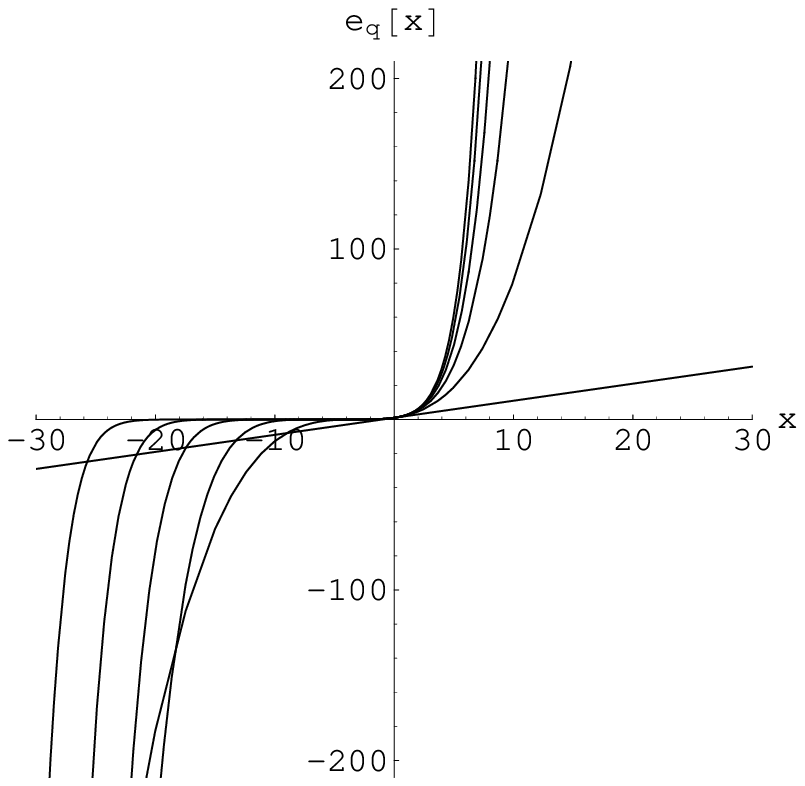}
\includegraphics[scale=0.84]{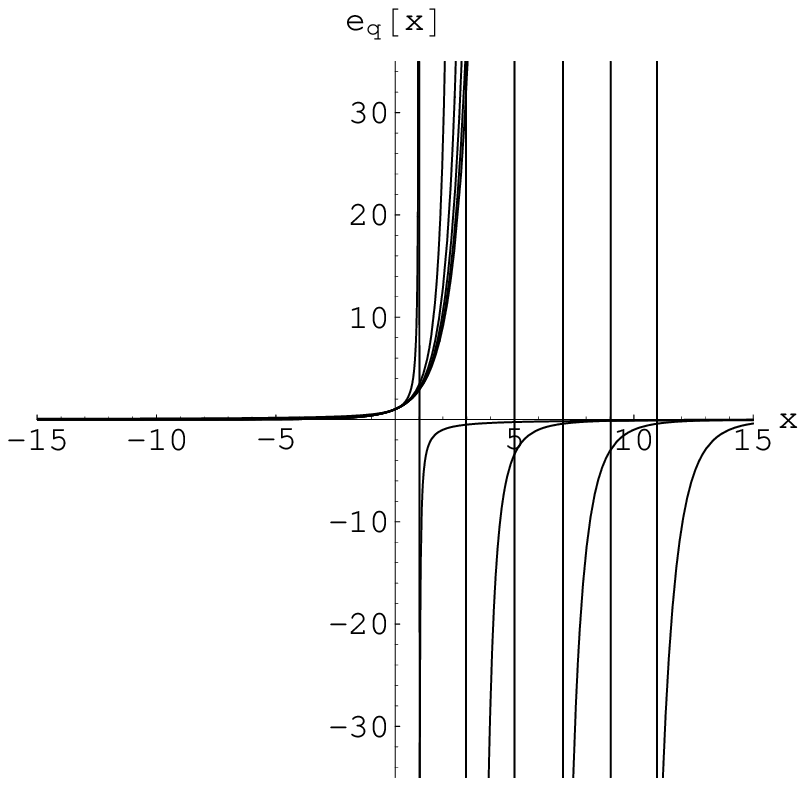}
\end{center}

\hspace{7mm} \footnotesize Fig.3 $q=\{0, \frac{2}{3}, \frac{4}{5}, \frac{6}{7}, \frac{8}{9},
\frac{10}{11}, \frac{12}{13}$\} \hspace{22mm} Fig.4
$q=\{\frac{12}{11},\frac{10}{9},\frac{8}{7},\frac{6}{5},\frac{4}{3}, 2$\}

\vspace{7mm}

\footnotesize Note: the $q<1$ and $q>1$ cases are displayed with different scales in order to
convey a qualitative understanding.

\normalsize

\end{appendix}

\end{document}